\documentclass
[nofootinbib,preprint,12pt,prd,groupedaddress,showpacs,eqsecnum,titlepage,showkeys]{revtex4}%
\usepackage{amsfonts}
\usepackage{amsmath}
\usepackage{amssymb}
\usepackage{graphicx}
\usepackage{latexsym}
\usepackage{color}%
\begin{document}
\title{Dilatonic effects near naked singularities}
\author{J.R. Morris}
\affiliation{Physics Dept., Indiana University Northwest, 3400 Broadway, Gary, Indiana
46408, USA}
\email{jmorris@iun.edu}

\begin{abstract}
Static spherically symmetric solutions of 4d Brans-Dicke theory include a set
of naked singularity solutions. Dilatonic effects near the naked singularities
result in either a shielding or an antishielding effect from intruding massive
test particles. One result is that for a portion of the solution parameter
space, no communication between the singularity and a distant observer is
possible via massive particle exchanges. Kaluza-Klein gravity is considered as
a special case.

\end{abstract}

\pacs{04.50.Kd, 04.20.Jb, 04.50.-h}
\keywords{Brans-Dicke theory, naked singularity, Kaluza-Klein gravity, exact solutions,
dilaton gravity}\maketitle

\section{\bigskip Introduction}

The static spherically symmetric vacuum solutions (see, for example,
\cite{Brans,X-Z,Cai-Myung}) of Brans-Dicke theory\cite{Brans-Dicke} include
the Schwarzschild solution, along with a set of solutions describing naked
singularities. The solutions of the 4-dimensional Brans-Dicke theory were
first constructed in the Jordan-Brans-Dicke frame by Brans\cite{Brans}.
However, a conformal transformation to the Einstein frame transforms the
matter-free Brans-Dicke theory into an Einstein theory of gravity with a
minimally coupled massless scalar field. The higher dimensional spherically
symmetric vacuum solutions of this theory in $D\geq4$ dimensions were provided
by Xanthopoulos and Zannias\cite{X-Z}, and include the 4-dimensional Brans
solutions, expressed in the Einstein frame. These two parameter $D-$%
dimensional solutions were shown to correspond to naked singularities for all
cases except for the case corresponding to the $D-$dimensional Schwarzschild solution.

\bigskip

Cai and Myung\cite{Cai-Myung} have also studied such solutions for the case of
$D\geq4$ dimensional Brans-Dicke theory in both the Jordan frame and the
Einstein frame, as well as examining solutions for the $D-$dimensional
Brans-Dicke-Maxwell theory. Again, for the case of neutral nonrotating
solutions the only \textquotedblleft black hole\textquotedblright\ solution is
the Schwarzschild solution with a constant dilaton, and the rest of the
solutions, describing naked singularities, have attendant nontrivial dilaton
fields. A 4d dilaton field $\tilde{\phi}(x^{\mu})$ in the Brans-Dicke theory,
and its spherically symmetric static solutions, can be connected in a
straightforward way with a scale factor $b(x^{\mu})$ of isotropically,
toroidally compactified extra spatial dimensions in a higher dimensional
theory of Einstein gravity. Specifically, $(4+n)-$dimensional pure Einstein
gravity, when dimensionally reduced to 4 dimensions, can take the form of a 4d
Brans-Dicke theory, with a Brans-Dicke parameter $\omega_{BD}$ that depends
upon the number $n$ of compactified dimensions and a Brans-Dicke field
$\tilde{\phi}$ that is related to the scale factor $b$ of the $n$ extra
dimensions. In this case of Kaluza-Klein gravity, a variation in the dilaton
field $\tilde{\phi}(x)$ is associated with a variation in the extra
dimensional scale factor $b(x)$. In the Einstein frame of the 4d action which
includes matter, there is a dilaton coupling to matter which results in
particle masses that depend on $b(x)$.

\bigskip

In \cite{NDL-JM} a study was made of the reflection and transmission of
massless and massive particles through a \textquotedblleft
wall\textquotedblright\ of varying $\tilde{\phi}$ or $b$ in the Einstein frame
of a Kaluza-Klein theory with $n$ extra dimensions that are isotropically
compactified. This study assumed a flat 4d spacetime background, and the
qualitative nature of the results agree with those of previous studies with
nondilatonic walls\cite{VSbook,Everett,EDW1,EDW2}. However, caution must be
exercised in applying these results to regions of strong spacetime curvature,
since there are gravitational redshifting effects to consider that may modify
the energy dependent reflection coefficient $\mathcal{R}(\omega)$ in such regions.

\bigskip

Problems associated with a calculation of reflection and transmission
coefficients of massive particles in strong gravitational fields are
sidetracked here through an approach that focuses upon the kinematics of
massive particles propagating in a dilaton-gravity background of the type
described by the neutral static spherically symmetric solutions studied by
Brans\cite{Brans}, Xanthopoulos and Zannias\cite{X-Z}, and Cai and
Myung\cite{Cai-Myung}. This approach allows an inference of some basic
physical effects on massive particles of the \textquotedblleft dilaton
cloud\textquotedblright\ surrounding the Brans-Dicke naked singularities. This
dilaton cloud is described by the exact analytical solution for the
Brans-Dicke scalar field.

\bigskip

Therefore, the approach presented here simply relies on kinematical properties
of massive test particles propagating in a dilatonic field of a neutral
nonrotating naked singularity, considered in the Einstein frame. Regions where
a particle is kinematically allowed to propagate are defined, and some general
comments are offered concerning purely radial motion. (Also see
ref.\cite{Bhamra}, where geodesic motion was considered in the Jordan frame,
using nonisotropic coordinates.) One set of solutions is associated with an
attractive dilatonic force on test particles, and another set of solutions
exerts a repulsive dilatonic force on test particles.

\bigskip

The results that are obtained suggest that, for a portion of the solution
parameter space describing naked singularities surrounded by a repulsive
dilaton cloud, test particles with nonzero mass infalling from radial
infinity, regardless of the particle energy, can not reach the singularity and
are reflected back out. That is, for a range of solution parameters, certain
regions close to the singularity become kinematically forbidden. Consequently,
for these solutions no communication between the singularity and a distant
observer via massive particle exchanges is possible. (This situation does not
appear to have been pointed out in ref.\cite{Bhamra}.) These results are also
expressed in terms of the number $n$ of compactified extra dimensions for the
special case of Kaluza-Klein gravity.

\bigskip

\section{Jordan and Einstein frame representations}

\subsection{Brans-Dicke gravity in 4d}

The action for four dimensional Brans-Dicke theory is given by%
\begin{equation}
S=\frac{1}{16\pi}\int d^{4}x\sqrt{\tilde{g}}\left\{  \tilde{\phi}\tilde
{R}+\frac{\omega_{BD}}{\tilde{\phi}}\tilde{g}^{\mu\nu}\partial_{\mu}%
\tilde{\phi}\partial_{\nu}\tilde{\phi}\right\}  +S_{m}(\tilde{g}_{\mu\nu})
\label{a1}%
\end{equation}

where $S_{m}$ is the matter action, Newton's constant is set to unity, $G=1$,
and $\tilde{g}=|\det(\tilde{g}_{\mu\nu})|$. This action is expressed in the
Jordan-Brans-Dicke conformal frame, or just the Jordan frame, for short. The
metric in this frame is $\tilde{g}_{\mu\nu}$ and the scalar curvature
$\tilde{R}$ is built from this metric. (A metric with signature $(+,-,-,-)$ is
used.) The dimensionless constant $\omega_{BD}$ is the Brans-Dicke parameter,
and the matter action $S_{m}(\tilde{g}_{\mu\nu})$ is constructed using the
Jordan frame metric $\tilde{g}_{\mu\nu}$. For a test particle of constant mass
$m_{0}$ in the Jordan frame, $S\propto\int m_{0}d\tilde{s}$, where $d\tilde
{s}=\sqrt{\tilde{g}_{\mu\nu}dx^{\mu}dx^{\nu}}$.

\bigskip

A conformal transformation to the Einstein frame is given by\cite{Cai-Myung}%
\begin{equation}
g_{\mu\nu}=\tilde{\phi}\tilde{g}_{\mu\nu},\ \ g^{\mu\nu}=\tilde{\phi}%
^{-1}\tilde{g}^{\mu\nu},\ \ \ \sqrt{g}=\tilde{\phi}^{2}\sqrt{\tilde{g}%
},\ \ \ \phi=\sqrt{2a}\ln\tilde{\phi},\ \ \ a=\omega_{BD}+\frac{3}{2}
\label{a2}%
\end{equation}

and the action in the Einstein frame then takes the form%
\begin{equation}
S=\frac{1}{16\pi}\int d^{4}x\sqrt{g}\left\{  R+\frac{1}{2}g^{\mu\nu}%
\partial_{\mu}\phi\partial_{\nu}\phi\right\}  +S_{m}(\tilde{\phi}^{-1}%
g_{\mu\nu}) \label{a3}%
\end{equation}

where $R$ is built from $g_{\mu\nu}$ and Einstein gravity is coupled to a
massless scalar dilaton field $\phi$. Note, however, that from the Einstein
frame perspective, mass becomes position dependent in general, since we can
write%
\begin{equation}
S_{m}\propto\int m_{0}d\tilde{s}=\int m_{0}\tilde{\phi}^{-1/2}ds=\int mds
\label{matter1}%
\end{equation}

where $d\tilde{s}=\tilde{\phi}^{-1/2}ds$ and $m$ is the mass in the Einstein
frame, given in terms of the dilaton field $\tilde{\phi}$ by%
\begin{equation}
m=\tilde{\phi}^{-1/2}m_{0} \label{mass1}%
\end{equation}

Therefore, a particle having a constant mass $m_{0}$ in the Jordan frame will
have a mass $m=\tilde{\phi}^{-1/2}m_{0}$ in the Einstein frame\cite{Dicke62}.

\bigskip

\subsection{Kaluza-Klein gravity}

Consider an action describing pure Einstein gravity coupled to matter in
$D=4+n$ dimensions,%
\begin{equation}
S_{D}=\int d^{D}x\sqrt{\tilde{g}_{D}}\left\{  \frac{1}{2\kappa_{D}^{2}}\left[
\tilde{R}_{D}[\tilde{g}_{MN}]-2\Lambda\right]  +\mathcal{\tilde{L}}%
_{D}\right\}  \label{S_D}%
\end{equation}

in a spacetime described by%
\begin{equation}
d\tilde{s}_{D}^{2}=\tilde{g}_{MN}dx^{M}dx^{N}=\tilde{g}_{\mu\nu}(x)dx^{\mu
}dx^{\nu}-\ b^{2}(x)\gamma_{mn}(y)dy^{m}dy^{n} \label{metric}%
\end{equation}

with $x^{M}=(x^{\mu},y^{m})$, $M,N=0,1,2,3,\cdot\cdot\cdot,D-1$, and
$\tilde{g}_{D}=|\det\tilde{g}_{MN}|$. There are $n$ extra compact space
coordinates labeled by $y^{m}$, and the metric for the extra dimensions is
$\tilde{g}_{mn}(x,y)=-b^{2}(x)\gamma_{mn}(y)$ with $\gamma_{mn}(y)$ describing
the geometry of the compact space, and the isotropic extra dimensional scale
factor $b(x^{\mu})$ is assumed to be $y$ independent. (The metric has
signature $(+,-,\cdot\cdot\cdot,-)$.) The factor $\kappa_{D}^{2}=8\pi G_{D}$
for the $D$ dimensional spacetime is related to the corresponding 4d one by
$\kappa_{D}^{2}=8\pi G_{D}=V_{y}\kappa^{2}=V_{y}(8\pi G)$ where $V_{y}$ is the
coordinate volume of the extra dimensional space and $\kappa^{2}=8\pi G=8\pi$
($G=1$) is the inverse of the 4d reduced Planck mass. $\mathcal{\tilde{L}}%
_{D}$ and $\Lambda$ are the matter Lagrangian and cosmological constant,
respectively, in the $D$ dimensional space.

\bigskip

Borrowing results used in\cite{CGHW} and\cite{NDL-JM} to express the action as
an effective 4d action, an integration over the $y$ coordinates leaves (sign
errors appearing in eq.(7) of ref.\cite{NDL-JM}, having no effect there, are
corrected here)
\begin{equation}%
\begin{array}
[c]{ll}%
S & =%
%TCIMACRO{\dint }%
%BeginExpansion
{\displaystyle\int}
%EndExpansion
d^{4}x\sqrt{\tilde{g}}\left\{  \dfrac{1}{2\kappa^{2}}[b^{n}\tilde{R}[\tilde
{g}_{\mu\nu}]+2nb^{n-1}\tilde{g}^{\mu\nu}\tilde{\nabla}_{\mu}\tilde{\nabla
}_{\nu}b+n(n-1)b^{n-2}\tilde{g}^{\mu\nu}(\tilde{\nabla}_{\mu}b)(\tilde{\nabla
}_{\nu}b)\right. \\
& \left.  +n(n-1)kb^{n-2}]+b^{n}\left[  \mathcal{L}_{D}-\dfrac{\Lambda}%
{\kappa^{2}}\right]  \right\}
\end{array}
\label{a4}%
\end{equation}

where $\mathcal{L}_{D}=V_{y}\mathcal{\tilde{L}}_{D}$ is a normalized
Lagrangian and $\tilde{R}[\tilde{g}_{\mu\nu}]$ is the Ricci scalar built from
$\tilde{g}_{\mu\nu}$, etc. (See \cite{CGHW} and\cite{NDL-JM} for details.) The
metric $\tilde{g}_{\mu\nu}$ then acts as a 4d Jordan frame metric. The
constant $n(n-1)k=\tilde{R}[\gamma_{mn}]$ gives the curvature of the internal
space, which we will set equal to zero. Dropping a total divergence, this
action can be rewritten as%
\begin{equation}%
\begin{array}
[c]{ll}%
S & =%
%TCIMACRO{\dint }%
%BeginExpansion
{\displaystyle\int}
%EndExpansion
d^{4}x\sqrt{\tilde{g}}\left\{  \dfrac{1}{2\kappa^{2}}[b^{n}\tilde{R}[\tilde
{g}_{\mu\nu}]-n(n-1)b^{n-2}\tilde{g}^{\mu\nu}(\tilde{\nabla}_{\mu}%
b)(\tilde{\nabla}_{\nu}b)\right. \\
& \left.  +n(n-1)kb^{n-2}]+b^{n}\left[  \mathcal{L}_{D}-\dfrac{\Lambda}%
{\kappa^{2}}\right]  \right\}
\end{array}
\label{a4a}%
\end{equation}

Upon setting $\tilde{R}[\gamma_{mn}]=n(n-1)k=0$ (toroidal compactification,
for example) and $\Lambda=0$, this assumes the form of (\ref{a1}) with the
identifications%
\begin{equation}
\tilde{\phi}=b^{n},\ \ \ \ \omega_{BD}=-1+\frac{1}{n},\ \ \ \mathcal{\tilde
{L}}_{m}=b^{n}\mathcal{L}_{D} \label{a5}%
\end{equation}
\bigskip Use of the conformal transformation of (\ref{a2}) allows the action
to be represented in the Einstein frame, as given by (\ref{a3}), with
$\phi=\sqrt{2a}\ln\tilde{\phi},\ \ \ a=\omega_{BD}+\frac{3}{2}$ (which gives
$a=\frac{n+2}{2n}$ for the Kaluza-Klein case), and an effective 4d matter
Lagrangian density $\mathcal{L}_{m}=b^{-n}\mathcal{L}_{D}$.

\bigskip

The 4d Einstein frame spacetime is described by $ds^{2}=g_{\mu\nu}(x)dx^{\mu
}dx^{\nu}$ and is related to the 4d Jordan frame spacetime of $d\tilde{s}%
^{2}=\tilde{g}_{\mu\nu}(x)dx^{\mu}dx^{\nu}$ by the conformal transformation
$\tilde{g}_{\mu\nu}=\tilde{\phi}^{-1}g_{\mu\nu}$. The two line elements are
related by $ds^{2}=\tilde{\phi}d\tilde{s}^{2}$. We take the classical action
for a test particle to be $S\propto\int m_{0}d\tilde{s}_{D}$ and assume the
particle worldline to be independent of the coordinates $y^{m}$, so that
$dy^{m}=0$ along the worldline. Then along the worldline $d\tilde{s}%
_{D}=d\tilde{s}=\tilde{\phi}^{-1/2}ds$ so that $S\propto\int m_{0}d\tilde
{s}_{D}=\int m_{0}d\tilde{s}=\int mds$, where%
\begin{equation}
m=\tilde{\phi}^{-1/2}m_{0}=b^{-\frac{n}{2}}m_{0} \label{mass2}%
\end{equation}

In the Jordan frame particles have constant, $\tilde{\phi}$ independent masses
and follow geodesics, whereas in the Einstein frame masses become $\tilde
{\phi}$ dependent and paths are generally not geodesics due to the $x^{\mu}$
dependence of the particle mass\cite{Dicke62}. The 4d matter fields in the
field theoretic Lagrangian density $\mathcal{L}_{m}$ can be rescaled in the
Einstein frame, but masses pick up a $\tilde{\phi}$ dependence (therefore a
$b$ dependence for the Kaluza-Klein case).

\bigskip

\bigskip

\section{Spherically symmetric vacuum solutions}

The vacuum solutions of the field equations obtained from the Einstein frame
action of (\ref{a3}), as well as higher dimensional generalizations, were
obtained by Xanthopoulos and Zannias\cite{X-Z}. These were also presented by
Cai and Myung\cite{Cai-Myung}. The static neutral solutions, with isotropic
coordinates, are presented here for the special 4d case:%
\begin{equation}
ds^{2}=e^{f}dt^{2}-e^{-h}(dr^{2}+r^{2}d\Omega^{2})\smallskip\label{a7a}%
\end{equation}

\begin{equation}
e^{f}=g_{00}=\xi^{2\gamma}\smallskip;\ \ \ \ \xi=\left(  \dfrac{r-r_{0}%
}{r+r_{0}}\right)  \label{a7b}%
\end{equation}

\begin{equation}
e^{-h}=|g_{rr}|=\left(  1-\dfrac{r_{0}^{2}}{r^{2}}\right)  ^{2}\xi^{-2\gamma
}=e^{-f}\left(  1-\dfrac{r_{0}^{2}}{r^{2}}\right)  ^{2} \label{a7c}%
\end{equation}

\begin{subequations}
\label{a7d}%
\begin{align}
\phi &  =\pm\tilde{\gamma}\ln\xi=\sqrt{2a}\ln\tilde{\phi}\smallskip
;\label{a7d1}\\
\tilde{\phi} &  =\xi^{\Gamma}\label{a7d2}%
\end{align}

where $r_{0}$ and $\gamma$ are integration constants ($r_{0}>0$), and we have
defined%
\end{subequations}
\begin{equation}
\xi=\left(  \frac{r-r_{0}}{r+r_{0}}\right)  \leq1,\ \ \ \tilde{\gamma
}=[4(1-\gamma^{2})]^{1/2},\ \ \ \ \Gamma=\pm\frac{\tilde{\gamma}}{\sqrt{2a}%
}=\pm\left[  \frac{2}{a}(1-\gamma^{2})\right]  ^{1/2} \label{a8}%
\end{equation}
These are the Einstein frame fields and solutions, with $0\leq\gamma\leq1$ for
the description of physical (nonegative ADM mass) solutions. There is a naked
singularity at $r=r_{0}$ where $R=g^{\mu\nu}R_{\mu\nu}\rightarrow\infty$
unless $\gamma=1$ and $\phi=0$ (the Schwarzschild solution).

\bigskip

\textit{Note}: In the set of vacuum solutions presented in ref.\cite{X-Z},
only the solution with the $+$ sign in (\ref{a7d1}), i.e., $\phi
=+\tilde{\gamma}\ln\xi$, is presented. However, the second solution
$\phi=-\tilde{\gamma}\ln\xi$ is seen to exist due to the invariance of the
action and equations of motion (EoM) under the transformations $g_{\mu\nu
}\rightarrow g_{\mu\nu}$, $\phi\rightarrow-\phi$. Thus if $\phi$ is a solution
to the EoM, then so is $-\phi$ (see, for example, refs.\cite{Cai-Myung}
and\cite{W-R}). Therefore $\phi$ can be positive or negative, and the
Brans-Dicke scalar $\tilde{\phi}=\xi^{\Gamma}=\xi^{\pm|\Gamma|}$ can be either
a decreasing or an increasing function of $r$ and $\xi$.

\bigskip

For the case of Kaluza-Klein gravity in 4d, (\ref{a5}) gives $\tilde{\phi
}=b^{n}$ so that $b=\xi^{\Gamma/n}$ with $\omega_{BD}=(1-n)/n$ and
$a=(n+2)/2n$. In this case $\Gamma$ takes a value%
\begin{equation}
\Gamma=\pm2\left[  \left(  \frac{n}{n+2}\right)  (1-\gamma^{2})\right]  ^{1/2}
\label{a9}%
\end{equation}

This allows the scale factor to either shrink to zero or to blow up as
$r\rightarrow r_{0}$,%
\begin{equation}
b\rightarrow\left\{
\begin{array}
[c]{cc}%
0, & \Gamma>0\\
\infty, & \Gamma<0\\
1, & \Gamma=0
\end{array}
\right\}  \text{\ \ \ as\ \ \ }r\rightarrow r_{0} \label{a9a}%
\end{equation}

These possibilities were pointed out by Davidson and Owen\cite{D-O} for the
case of one extra dimension, $n=1$. The case $\gamma=1$, $\Gamma=0$
corresponds to the Schwarzschild solution, with $\phi=0$, $\tilde{\phi}=b=1$.

\section{Kinematical constraint for massive particle}

Now attention is focused on a spinless test particle of arbitrary nonzero mass
$m(r)$ propagating in the spacetime of (\ref{a7a}). The particle mass $m$ is
generally position dependent, due to the dilaton field, $m=\tilde{\phi}%
^{-1/2}m_{0}$. We will obtain kinematical constraints on the allowed and
forbidden regions of particle propagation by considering a classical
\textquotedblleft geodesic\textquotedblright\ approach.

\subsection{Geodesic Approach}

Consider now a neutral test particle moving under the influence of the
gravitational and dilatonic fields present in the Einstein frame. In this
frame, again, $m(r)$ varies with radial position $r$ of the particle. This
gives rise to a dilatonic force correction to the geodesic
equation\cite{Dicke62} in the Einstein frame,%
\begin{equation}
\frac{d}{ds}\left(  mg_{\mu\nu}u^{\nu}\right)  -\frac{1}{2}m(\partial_{\mu
}g_{\alpha\beta})u^{\alpha}u^{\beta}-\partial_{\mu}m=0 \label{e19}%
\end{equation}

where $u^{\mu}=dx^{\mu}/ds$ and the $\partial_{\mu}m$ term arises due to the
variability of $m$. \ The velocities $u^{\mu}$ satisfy the constraint
$g_{\mu\nu}u^{\mu}u^{\nu}=1$. Focusing upon a particle with a purely radial
trajectory with $\theta,\phi$ held constant, we have the components
$u^{0}=u^{t}$ and $u^{1}=u^{r}$ being nonvanishing, in general. The vacuum
solutions are described by a time independent, diagonal metric and dilaton,
$\partial_{0}g_{\mu\nu}=0$, $\partial_{0}m=0$, and the geodesic equations for
$u^{0}$ and $u^{r}$ are given by
\begin{align}
\frac{d}{ds}(m\ u_{0})-\frac{1}{2}m(\partial_{0}g_{\alpha\beta})u^{\alpha
}u^{\beta}-\partial_{0}m  &  =0\label{e20a}\\
\frac{d}{ds}(m\ u_{r})-\frac{1}{2}m(\partial_{r}g_{\alpha\beta})u^{\alpha
}u^{\beta}-\partial_{r}m  &  =0 \label{e20b}%
\end{align}

These reduce to
\begin{align}
&  \frac{d}{ds}(m\ u_{0})=\frac{d}{ds}(mg_{00}\ u^{0})=0\label{e21a}\\
&  \frac{d}{ds}(mg_{rr}u^{r})-\frac{1}{2}m\left[  (\partial_{r}g_{00}%
)(u^{0})^{2}+(\partial_{r}g_{rr})(u^{r})^{2}\right]  -\partial_{r}m=0
\label{e21b}%
\end{align}

The first equation gives%
\begin{equation}
u^{0}=\frac{\alpha}{mg_{00}} \label{e22}%
\end{equation}

where $\alpha=m\ u_{0}=p_{0}$ is a constant. The constraint $u_{\mu}u^{\mu
}=g_{00}(u^{0})^{2}+g_{rr}(u^{r})^{2}=1$ then produces
\begin{equation}
(u^{r})^{2}=\left(  \frac{dr}{ds}\right)  ^{2}=\frac{1}{g_{rr}}\left[
1-\frac{\alpha^{2}}{m^{2}g_{00}}\right]  \label{e23}%
\end{equation}

For $r>r_{0}$ where $g_{rr}<0$ is finite, the above constraint requires that
$\ $%
\begin{equation}
\ m^{2}g_{00}=m_{0}^{2}\tilde{\phi}^{-1}g_{00}\leq\alpha^{2} \label{e23a}%
\end{equation}
in regions where the particle is kinematically allowed to propagate, with
$(u^{r})^{2}\geq0$. Turning points of the radial motion are located where
$(u^{r})^{2}=0$.

\bigskip

Asymptotically, we have $g_{00}\rightarrow1$, $g_{rr}\rightarrow-1$, and
therefore at $r=\infty$, $u_{\mu}u^{\mu}=(u_{\infty}^{0})^{2}-(u_{\infty}%
^{r})^{2}=1$. The proper time of the particle moving with radial speed $v$ at
$r=\infty$ is given by $ds=d\tau=\gamma_{rel}^{-1}dt$, so that $u^{0}%
=dt/ds=\gamma_{rel}=1/\sqrt{1-v^{2}}$. (Here $\gamma_{rel}$ is the ordinary
special relativistic gamma factor, not to be confused with the solution
parameter $\gamma$.) Therefore%
\begin{equation}
\left(  u_{\infty}^{r}\right)  ^{2}=\gamma_{rel}^{2}-1 \label{e24}%
\end{equation}

Now using the definition of $\alpha$,%
\begin{equation}
\alpha=mg_{00}u^{0}=\left(  mg_{00}u^{0}\right)  |_{\infty}=\gamma_{rel}%
m_{0}=E_{0}=\omega_{0} \label{e25}%
\end{equation}

With this identification of $\alpha=\omega_{0}$ the kinematical constraint in
(\ref{e23a}) becomes $\left(  m^{2}/\omega_{0}^{2}\right)  g_{00}\leq1$, or%
\begin{equation}
\frac{m^{2}}{\omega^{2}}=\frac{m_{0}^{2}}{\omega_{0}^{2}}(\tilde{\phi}%
^{-1}g_{00})\leq1\text{\ \ \ (kinematically allowed,\ }\omega=\frac{\omega
_{0}}{\sqrt{g_{00}}}\text{)} \label{e26}%
\end{equation}

\subsection{Radial motion}

We can define an energy parameter $\mathcal{E}=E_{0}/m_{0}=\omega_{0}/m_{0}$,
which is the asymptotic energy per unit mass value. Eq. (\ref{e23}) can then
be expressed as%
\begin{equation}
(u^{r})^{2}=\left(  \frac{dr}{ds}\right)  ^{2}=\frac{1}{|g_{rr}|}\left[
\mathcal{E}^{2}\frac{\tilde{\phi}}{g_{00}}-1\right]  \label{26u}%
\end{equation}
The kinematically allowed region where the particle can propagate can then be
written as%
\begin{equation}
\mathcal{E}^{2}\frac{\tilde{\phi}}{g_{00}}=\mathcal{E}^{2}\left(
\frac{r-r_{0}}{r+r_{0}}\right)  ^{-\left(  2\gamma-\Gamma\right)
}=\mathcal{E}^{2}\xi^{-\left(  2\gamma-\Gamma\right)  }\geq
1\text{\ \ \ \ \ \ \ (kinematically allowed)} \label{e26a}%
\end{equation}

Turning points are located where $(u^{r})^{2}=0$, or $\mathcal{E}^{2}%
\frac{\tilde{\phi}}{g_{00}}=\mathcal{E}^{2}\xi^{-\left(  2\gamma
-\Gamma\right)  }=1$. A particle can escape to $r\rightarrow\infty$
($\xi\rightarrow1$) if $\mathcal{E}>1$. However, if $2\gamma-\Gamma\geq0$ but
$\mathcal{E}<1,$ $\xi=1$ is not kinematically allowed, and the particle will
be gravitationally trapped. For $\mathcal{E}>1$ and $2\gamma-\Gamma\geq0$,
there are no turning points. In this case a particle can plunge inward from
radial infinity all the way to the singularity, and a particle ejected from
the singularity region can escape to infinity. Note that the condition
$2\gamma-\Gamma\geq0$ is obtained for all $\Gamma\leq0$ and for positive
values of $\Gamma$ for which $\Gamma\leq2\gamma$.

\bigskip

If, on the other hand, $2\gamma-\Gamma<0$, then (\ref{e26a}) implies that
$1>\xi^{|2\gamma-\Gamma|}\geq1/\mathcal{E}^{2}$, and since $\xi<1$ we must
require $\mathcal{E}>1$, and the particle cannot be gravitationally trapped in
this case, and we have $\xi\in\lbrack\xi_{\min},1)$ and $r\in\lbrack r_{\min
},\infty)$, where $\xi_{\min}=\mathcal{E}^{-2/|2\gamma-\Gamma|}$ locates the
turning point where $r=r_{\min}$ and $u^{r}=0$. The condition $2\gamma
-\Gamma<0$ requires positive values of $\Gamma$, with $\Gamma>2\gamma$.

\section{Kinematically allowed and forbidden regions}

The constraint (\ref{e26}) or (\ref{e26a}) for kinematically allowed regions
is $m^{2}/\omega^{2}\leq1$ and the kinematically forbidden region has
$m^{2}/\omega^{2}>1$. \ The condition (\ref{e26a}) for kinematically allowed
regions is%
\begin{equation}
\frac{m^{2}}{\omega^{2}}=\frac{m_{0}^{2}}{\omega_{0}^{2}}(\tilde{\phi}%
^{-1}g_{00})=\frac{m_{0}^{2}}{\omega_{0}^{2}}\text{\ }\left(  \frac{r-r_{0}%
}{r+r_{0}}\right)  ^{2\gamma-\Gamma}=\frac{\xi^{2\gamma-\Gamma}}%
{\mathcal{E}^{2}}\leq1\ \text{\ \ \ (kinematically allowed)} \label{e28}%
\end{equation}

As $r\rightarrow r_{0}$ ($\xi\rightarrow0$),
\begin{equation}
\frac{m^{2}}{\omega^{2}}\rightarrow\left\{
\begin{array}
[c]{llll}%
0, & \text{if }2\gamma-\Gamma>0: & r\rightarrow r_{0}\ \ \text{is allowed,} &
2\gamma>\Gamma\\
\infty, & \text{if }2\gamma-\Gamma<0: & r\rightarrow r_{0}\ \ \text{is
forbidden,} & 2\gamma<\Gamma\\
\mathcal{E}^{-2}, & \text{if }2\gamma-\Gamma=0: & r\rightarrow r_{0}%
\ \ \text{is allowed,} & 2\gamma=\Gamma
\end{array}
\right\}  \label{e29}%
\end{equation}

The particle can travel all the way to $r_{0}$ with $m/\omega\leq1$ if
$2\gamma-\Gamma\geq0$, but it cannot reach $r=r_{0}$ for $2\gamma-\Gamma<0$,
i.e. if $2\gamma<\Gamma$. For the case $2\gamma=\Gamma$, we must have
$\mathcal{E}\geq1$.

\bigskip

Now, for the case $\Gamma>2\gamma\geq0$ we have $\frac{2\gamma}{\Gamma}\geq0$.
Then the singularity $r=r_{0}$ is forbidden only for $\frac{2\gamma}{\Gamma
}<1$, and this leads to the parameter constraint $\gamma=\sqrt{\frac
{(2\gamma/\Gamma)^{2}}{(2\gamma/\Gamma)^{2}+2a}}<\frac{1}{\sqrt{1+2a}}$: i.e.,
for
\begin{equation}
\Gamma>0,\ \ \ \ \ \gamma<\frac{1}{\sqrt{1+2a}}:\ \ \text{\ \ \ \ \ \ \ }%
r=r_{0}\text{ \ is forbidden} \label{e30}%
\end{equation}

The \textit{singularity is a forbidden region} for a vacuum solution with
$\Gamma>0$ and $\gamma<(1+2a)^{-1/2}$. \ For the case of Kaluza-Klein gravity,
$a=(n+2)/2n$ and (\ref{e30}) translates into%

\begin{equation}
\Gamma>0,\ \ \ \ \ \gamma<\frac{1}{\sqrt{2}}\sqrt{\frac{n}{n+1}}%
:\ \ \ \ \ \ \ r=r_{0}\text{ \ is forbidden} \label{e31}%
\end{equation}

The naked singularity solutions with positive $\Gamma$ and $\ \gamma
<(1+2a)^{-1/2}$ do not allow massive particles to reach the singular point
$r_{0}$, but the other solutions with $\gamma>(1+2a)^{-1/2}$ do. For the case
of Kaluza-Klein gravity, setting $n=1$ corresponds to $\gamma<\frac{1}{2}$,
while for large $n$ this approaches $\gamma<\frac{1}{\sqrt{2}}=.707$. In
either case, for Kaluza-Klein gravity there is a sizable portion of the
$\gamma\in\lbrack0,1]$ parameter space for which the naked singularity
solutions do not allow massive particles to propagate in the immediate
vicinity of the singularity.

\bigskip

On the other hand, in the context of a Brans-Dicke theory with a massless
scalar field $\tilde{\phi}$, which is subject to the constraint that
$\omega_{BD}\gg1$ ($a\gg1$), there is only a small portion of the $\gamma$
parameter space for which the singularity is untouchable by massive particles.
For example, the solar system constraint on $\omega_{BD}$ for a
\textit{massless} Brans-Dicke theory requires\cite{Bertotti} $\omega
_{BD}>40,000$. In this case, for $\omega_{BD}\sim a\gtrsim4\times10^{4}$,
(\ref{a7d2}) implies that $|\Gamma|\lesssim10^{-2}$, which lies very close to
the Schwarzschild limit $\Gamma=0$, and hence $\tilde{\phi}=\xi^{\Gamma}$ is
very slowly varying, and may make the $\gamma\neq1$ Brans-Dicke solutions
$\tilde{\phi}\neq1$ difficult to distinguish from the Schwarzschild solution
$\tilde{\phi}=1$. Note, however, that for $\gamma\neq1$, the Einstein frame
metric is distinct from the Schwarzschild metric (see (\ref{a7a}%
)-(\ref{a7c}).\bigskip

\subsection{\textbf{Minimal radius, }$\mathbf{r}_{c}$\textbf{:}\ \ }

We wish to find the minimal radius $r_{c}$ that a particle is allowed to
reach, or equivalently, the minimum radius from which the particle is
kinematically excluded, for the case where the singularity is kinematically
forbidden ($2\gamma<\Gamma$).\ To find $r_{c}$ write (\ref{e26a}) as
\begin{equation}
\xi=\left(  \frac{r-r_{0}}{r+r_{0}}\right)  \geq\left(  \frac{1}%
{\mathcal{E}^{2}}\right)  ^{\frac{1}{|2\gamma-\Gamma|}}=\xi_{c}%
>0;\ \ \ \ \ \ \ (2\gamma-\Gamma)<0 \label{e32}%
\end{equation}
where $\Gamma>2\gamma$ for the case where $r=r_{0}$ is kinematically
forbidden. This defines the kinematically excluded region with radius $r\leq
r_{c}$. The constant parameter $\xi_{c}=\xi_{c}(\mathcal{E})$ defined above is
positive and decreases with increasing particle energy parameter $\mathcal{E}%
$. However, for any finite value of $\mathcal{E}$ we have $\xi>0$, so that the
singularity is not reached. Setting $r=r_{c}$ for $\xi=\xi_{c},$ we have%
\begin{equation}
r_{c}=\left(  \frac{1+\xi_{c}}{1-\xi_{c}}\right)  r_{0} \label{e35}%
\end{equation}

and, from (\ref{e32}), $0<\xi_{c}<1$ since $\xi_{c}\leq\xi<1$. The
kinematically allowed region is that for which $r>r_{c}$. Restricting $\xi
_{c}$ to the range $\xi_{c}\in(0,1)$ implies that the particle energy
parameter $\mathcal{E}=\xi_{c}^{-\frac{|2\gamma-\Gamma|}{2}}$ for such a
solution with an untouchable singularity is restricted to the range
$\mathcal{E}\in(1,\infty)$. For a particle at rest ($\mathcal{E}=1$) at
$r\rightarrow\infty$ we see that $r_{c}\rightarrow\infty$, and the particle
will not gravitationally fall inward. A particle that is projected inward with
$\mathcal{E}>1$ will penetrate the dilaton cloud down to the radius
$r_{c}>r_{0}$. A particle cannot exist at $r<\infty$ with $\mathcal{E}<1$
without the application of an external force.

\subsection{Dilatonic acceleration}

The corrected geodesic equation (\ref{e19}) may be rewritten as%
\begin{equation}
\frac{du^{\nu}}{ds}=-\Gamma_{\alpha\beta}^{\nu}u^{\alpha}u^{\beta}%
+\partial_{\mu}(\ln m)(g^{\mu\nu}-u^{\mu}u^{\nu}) \label{e36}%
\end{equation}

The first term on the right hand side is the gravitational acceleration due to
the metric field $g_{\mu\nu}$, while the second term on the right hand side
represents the dilatonic acceleration, whose radial component is%
\begin{equation}
a_{r}^{dil}=\partial_{r}(\ln m)[g^{rr}-(u^{r})^{2}]=\frac{1}{2}\Gamma
\partial_{r}(\ln\xi)\left[  \frac{\mathcal{E}^{2}\xi^{\Gamma}}{|g_{rr}|g_{00}%
}\right]  \label{e37}%
\end{equation}

where use has been made of (\ref{mass1}) or (\ref{mass2}), (\ref{a7d2}), and
(\ref{26u}). Noting that the term $\partial_{r}(\ln\xi)$ is positive, as is
the term in square brackets, we see that the sign of the radial component of
the dilatonic acceleration $a_{r}^{dil}$ depends upon the sign of $\Gamma$.
For $\Gamma=0$ (the Schwarzschild solution) $a_{r}^{dil}=0$. For $\Gamma\neq
0$,%
\begin{equation}
a_{r}^{dil}\text{\ is }\left\{
\begin{array}
[c]{cc}%
>0\text{ for }\Gamma>0 & \text{(radial outward acceleration)}\\
<0\text{ for }\Gamma<0 & \text{(radial inward acceleration)}%
\end{array}
\right\}  \label{e38}%
\end{equation}

This corresponds to the fact that $m(r)\propto\tilde{\phi}^{-1/2}=\xi
^{-\Gamma/2}$ decreases radially outward for $\Gamma>0$ and $m(r)$ increases
radially outward for $\Gamma<0$, so that the constraint $p_{\mu}p^{\mu}%
=m^{2}(r)$ implies that the test particle is attracted to regions of lower
mass. The dilatonic repulsion ($\Gamma>0$) or dilatonic attraction ($\Gamma
<0$) of a test particle is therefore sensitive to the parameter $\Gamma$.

\section{Summary}

Static, spherically symmetric solutions of matter-free Brans-Dicke theory
describe a class of naked singularities. The effect of the dilaton cloud
(Brans-Dicke scalar field) on the radial motion of test particles has been
found to have either a shielding or antishielding effect, depending on the
values of the solution parameters. (Kaluza-Klein gravity has been examined as
a special case.) The Brans-Dicke scalar field (see (\ref{a7d})) $\tilde{\phi
}=\xi^{\Gamma}$ depends on the parameter $\Gamma$, which, in turn, depends
upon the solution parameter $\gamma\in\lbrack0,1]$ and the Brans-Dicke
parameter $\omega_{BD}$. The special case $\gamma=1$ yields the Schwarzschild
solution, for which $\Gamma=0$ and there is no dilatonic effect, with
$\tilde{\phi}=1$. However, naked singularity solutions with $\gamma\neq1$ have
$\Gamma\neq0$ with a radially varying dilaton field $\tilde{\phi}(r)$. In
spite of this, since $\Gamma$ depends inversely on $\sqrt{\omega_{BD}}$, a
very large parameter $\omega_{BD}\gg1$ can give rise to $\Gamma\approx0$ with
$\gamma\neq1$. In this case the dilaton field is slowly varying with possibly
negligible effects, though the Einstein frame metric of the spacetime depends
on $\gamma$ and deviates from the Schwarzschild case for $\gamma\neq1$.
Specifically, for $\gamma\neq1$, but $\Gamma\approx0$, the Brans-Dicke scalar
is nearly frozen at a value $\tilde{\phi}\approx1$, so that a test particle
essentially follows a geodesic (see eqs.(\ref{e36}) and (\ref{e37})), but not
a Schwarzschild one.

\bigskip

A kinematical constraint on a test particle of mass $m$ has been established,
describing kinematically allowed regions where the particle may exist. This
has been done in the Einstein frame where $m(r)\propto\tilde{\phi}^{-1/2}(r)$.
In the Einstein frame we make use of the (extended) set of
Xanthopoulos-Zannias solutions, for which the scalar field $\phi$ can take
positive or negative values, and hence $\tilde{\phi}$, which depends upon
$e^{\phi}$, can be either an increasing or decreasing function of $r$. (In the
Kaluza-Klein case this corresponds to the extra dimensional scale factor
$b(r)$ either vanishing or blowing up on the singularity.) The simple
kinematical constraint obtained allows a determination of relative values of
$\gamma$ and $\Gamma$ for which the naked singularity is kinematically
accessible to the particle.

\bigskip

We find that the singularity is inaccessible, i.e., forbidden, for the
parameters $\Gamma>2\gamma$. In this case there can be no communication via
massive particles between the singularity and a distant observer at infinity.
For the case of $\Gamma>0$ there is a dilatonic repulsion of the test
particle, and for $\Gamma<0$ there is a dilatonic attraction. The dilatonic
acceleration is sensitive to the parameter $\Gamma$, and for the case of
$\Gamma>2\gamma$ the closest distance of approach of the particle to the
singularity is found to depend upon the particle's energy parameter
$\mathcal{E}$ and the quantity $(\Gamma-2\gamma)$. For $\Gamma\leq2\gamma$ the
particle will be gravitationally trapped if $\mathcal{E}<1$, but can escape to
$r\rightarrow\infty$ if $\mathcal{E}>1$. On the other hand, for $\Gamma
>2\gamma$ the particle cannot be gravitationally trapped.

\bigskip

\textbf{Acknowledgement:}\ I thank an anonymous referee for useful comments.

\end{document}